\documentclass[conference]{IEEEtran}
\IEEEoverridecommandlockouts

\usepackage{cite}
\usepackage{amsmath,amssymb,amsfonts}
\usepackage{algorithmic,subcaption,hyperref}
\usepackage{graphicx,multirow}
\usepackage{textcomp}
\usepackage{booktabs}
\usepackage{siunitx}
\usepackage{makecell}
\usepackage{xcolor}
\def\BibTeX{{\rm B\kern-.05em{\sc i\kern-.025em b}\kern-.08em
    T\kern-.1667em\lower.7ex\hbox{E}\kern-.125emX}}
\begin{document}

\title{Learning Input-Channel Permutation Equivariance for Multi-Channel Source Separation: Reducing Bleeding in Small Music Ensembles
\thanks{This work was supported by “REPERTORIUM” Project under Grant Agreement 101095065. Horizon Europe. Cluster II. Culture, Creativity and
 Inclusive Society. Call HORIZON-CL2-2022-HERITAGE-01-02. \\ 
The authors wish to acknowledge CSC—IT Center for Science, Finland, for computational resources. }
}

\author{\begin{tabular}{c}Ruchi Pandey$^*$, Jaime Garcia-Martinez$^\dagger$, Pablo Cabañas-Molero$^\dagger$, David Diaz-Guerra$^*$,\\ Ricardo Falcón Pérez$^*$, Tuomas Virtanen$^*$, Julio J. Carabias-Orti$^\dagger$, and Pedro Vera-Candeas$^\dagger$\end{tabular} \\
$^*$ Audio Research Group, Tampere University, Tampere, Finland \\
$^\dagger$ Telecommunication Engineering Department, University of Jaen, Spain
}

\maketitle

\begin{abstract}
Microphone bleed is a persistent challenge in small ensembles and orchestral recordings, where close microphones intended for individual instruments also capture leakage from nearby sources. This overlap degrades track isolation and complicates mixing. This paper addresses the bleeding problem by making channel-permutation-equivariance a core learning principle. During training, we apply the same random permutation to the input microphone channels and their corresponding reference targets. This discourages reliance on fixed channel–instrument associations and improves robustness to changes in the recording setup and even in the recorded instruments. The proposed model is trained on synthetic ensembles with diverse simulated room acoustics and microphone placements, and evaluated on unseen simulated conditions and real URMP recordings. The results show that permutation-aware training consistently improves SDR and reduces bleeding under unseen conditions compared with non-permutation baselines. The findings highlight permutation-equivariance as a simple, data-centric strategy for robust debleeding and practical multi-channel source separation in music production workflows.
\end{abstract}


\section{Introduction}
Contrary to contemporary studio music, where instruments are often recorded one at a time, classical music is typically recorded with all instruments playing together in the same room. A main stereo microphone captures the full mixture and room reverberation, and close microphones are added to capture the individual instruments or sections and allow the mixing engineers to reinforce the weaker instruments at the main stereo microphone. In orchestral recordings, these close microphones are placed near every instrument section, and in small ensembles such as string quartets, they are often attached directly to each instrument. However, close microphones inevitably capture sound from neighboring sources, creating microphone bleed. This overlap degrades track isolation \cite{das2021, Terrell2011-wl} and complicates mixing, since increasing the amplitude of one instrument through its close mic also amplifies the leakage from others.

Reducing bleed between close microphones can be formulated as a music source separation (MSS) problem, a field recently transformed by deep learning methods \cite{sawata2021all,defossez2021hybrid,luo2023music}. However, these advances are based on the MUSDB18 setup \cite{rafii2017musdb18}, which focuses on single-channel or stereo separation of four stems: drums, bass, voices, and others. Multi-channel MSS is especially relevant for orchestral music, where multiple instruments of similar timbre interact in reverberant performance halls \cite{miron2016score}, and utilizing both spectral and spatial cues is crucial \cite{araki202530+}. Despite its importance, multichannel MSS has received little attention since deep learning became dominant. 

Existing multichannel approaches are more common in speech separation \cite{luo2020end,wang2021multi,tesch2023multi}, but these models are designed for microphone arrays rather than close-microphone setups. A related approach in \cite{wang2024mixture} uses close-mic signals for weak supervision in speech separation. However, its goal is array-based speech enhancement, not reducing bleed or cross-talk between close microphones. In \cite{francesc_direction_2023,garcia2025Ambi}, Higher-Order Ambisonics beamformers were used to generate virtual close microphones, with debleeding applied independently to each channel as a single-channel task. In contrast, this work formulates debleeding as a multi-channel source separation problem. We process all $P$ close-microphone signals jointly and estimate $P$ cleaned close-mic tracks, one per input channel. Unlike microphone-array separation, close-mic setups have one microphone per source and the dominant signal differs by channel, while residual bleed is spatially colored and varies with layout and room conditions. This motivates training separators that are robust to changes in the recording setup.

Supervised training requires isolated stems, which are rarely released due to copyright restrictions. Public datasets such as MUSDB18 \cite{rafii2017musdb18} exist, yet they are limited to stereo recordings of popular music. For classical instruments, synthetic datasets have recently been introduced \cite{sarkar2022ensembleset,garcia2025synthsod,dabike2024cadenza}, but models trained on synthetic data often struggle to generalize to real recordings \cite{garcia2025synthsod}. Although close microphones provide higher signal-to-distortion ratios (SDRs) than stereo mixtures, debleeding small ensembles remains difficult due to the lack of real ground-truth recordings for many instruments and the need to adapt to diverse acoustic conditions. 

To mitigate this synthetic-to-real gap, we enforce channel-permutation equivariance during training, reducing the reliance of the model on the timbral properties of the instruments and making it focus on the inter-microphone clues. We do this by applying the same random permutation to the input channels and their corresponding reference targets. Unlike Permutation Invariant Training (PIT), which addresses output–label assignment ambiguity, our outputs remain tied to the input channels and permutations are used only to improve generalization. We explicitly test generalization to unseen simulated rooms/layouts and to real URMP recordings \cite{li2018creating}, while noting that multichannel close-mic datasets with ground-truth debleeded references remain scarce.

The main contributions of this work are: (i) the simulation of close-microphone mixtures using synthetic instrument sources and simulated RIRs, (ii) the adaptation of the Hybrid Demucs architecture to map $P$ close-mic inputs to $P$ debleeded outputs, (iii) the enforcement of channel-permutation equivariance during training, and (iv) a comprehensive evaluation under different simulated recording setups including real audio sources. 

\section{Debleeding problem}
\label{sec:methodo}
In this section, we formulate \emph{debleeding} as a multichannel source separation problem with spatially colored cross–talk. Consider $P$ close microphones, each nominally assigned to one instrument. Let $x_i[n]$ denote the signal of the intended instrument at the $i^{\text{th}}$ microphone. For the recorded signals ${y_i[n]}_{i=1}^P$, the mixture model is defined as
\begin{equation}
y_i[n] = x_i[n] + \sum_{j\neq i} \alpha_{ij}[n] * x_j[n],
\end{equation}
where $\sum_{j\neq i} \alpha_{ij}[n] * x_j[n]$ is the bleed from other sources, $\alpha_{ij}[n]$ is the cross-channel acoustic impulse response, and $*$ denotes convolution. The objective of \emph{debleeding} is to recover clean estimates $\hat{x}_i[n] = [x_1[n],\ldots, x_P[n]]^\top$ from the multichannel observation $\mathbf{y}[n] = [y_1[n],\ldots, y_P[n]]^\top$, by suppressing the inter-instrument bleeding.
\section{Proposed method and Experimental Setup}
\label{sec:exp_setup}
\subsection{Learning channel-permutation equivariance}
We train the proposed model on synthetic data, where individual instrument signals can be rendered with controllable room acoustics and microphone layouts. A key challenge with such data is the risk of overfitting: models may memorize the timbral characteristics or the fixed spatial relationships of simulated instruments, which leads to poor generalization on real recordings. To address this, we introduce a channel-permutation strategy during training. The input microphone channels and their corresponding reference outputs are randomly permuted for every dataset elements during training. Formally, for any permutation $\pi$ of the $P$ channels, the training objective encourages $f(\pi(y)) \approx \pi(f(y))$; i.e., equivariance to channel permutations. Across mini-batches and epochs, the same instrument may appear in different channel indices. This prevents the model from relying on a fixed channel–instrument assignment and encourages a permutation-equivariant mapping. The model must debleed whichever instrument dominates a channel using inter-channel cues, rather than memorizing timbre or channel order. This randomization also restricts the model from exploiting static inter-channel relationships in the synthetic data. Instead, it encourages the model to infer inter-channel acoustic relationships directly from the input during inference. 

In literature, PIT is explored in related domains such as speech separation \cite{YuKT017}, sound event localization and detection \cite{CaoIKZWP20}, and multi-object tracking \cite{abs-1907-12887}. PIT typically resolves the output–label ambiguity by considering all possible predictions and ground truth permutations when computing the loss. In contrast, our setting has no such ambiguity because each output channel corresponds to an input microphone channel. We therefore apply the same permutation to inputs and targets to enforce permutation equivariance and robustness to channel reordering.
\subsection{Multichannel separation model}
We use the Hybrid Demucs architecture as the basis of our debleeding model. Hybrid Demucs is a dual-domain extension of Demucs \cite{defossez2021hybrid}. The model combines waveform and spectrogram modeling for both efficiency and separation quality. The model consists of two parallel branches. The temporal branch operates directly on raw waveforms, while the spectral branch processes STFT features (window size 4096, hop size 1024). Frequency-axis convolutions are applied in the spectral branch to produce time-aligned features that match the temporal encoder. The outputs of both branches are summed and passed through shared encoder–decoder layers with ×2 time downsampling. Finally, two domain-specific U-Nets, each with independent skip connections, decode the temporal and spectral streams. The final estimates are obtained by summing the temporal output with the inverse STFT of the spectral output. 

For the debleeding task, we adapt Hybrid Demucs from its original design for single stereo input and four stereo stems to a multichannel setting. The adapted model takes $P$ monaural close-microphone channels as inputs and outputs $P$ monaural estimates, enabling channel-wise debleeding while benefiting from joint multichannel context. 

We modify the input and output layers to operate on $P$ channels while keeping the core encoder–decoder blocks unchanged. We also compare two input representations for the frequency branch: (i) magnitude spectrograms and (ii) concatenated real and imaginary parts of the STFT. All models are trained with a time-domain L1 loss until the validation loss converges, which occurred after approximately 100 epochs in all considered cases. In this study, $P$ is fixed to $5$, which is the maximum ensemble size in our dataset.
\section{Evaluation}
\label{sec:results}
\subsection{Datasets}
We generate the training data from the SynthSOD stems, focusing on small string ensembles. Specifically, we select 106 music pieces, including Violin 1, Violin 2, Viola, Cello, and Bass stems. From these, 70 pieces are used for training, 18 for validation, and 18 for testing. In addition, we extract two subsets from the URMP dataset: one containing only string pieces and another containing non-string pieces.

To simulate microphone bleed under controlled conditions, we use a custom software framework built on top of PyRoomAcoustics’ ShoeBox model \cite{Scheibler_2018}. This framework allows us to simulate multiple recording environments characterized by their room dimensions and reverberation times (RT60). To make it realistic, we consider these environments on acoustical properties reported for real orchestral spaces in \cite{Adams2019}. Table \ref{tab:room_des} summarizes the specific room configurations. Rooms R1–R4 are used for training, while Rooms R5 and R6 are reserved for validation and testing, respectively.

\begin{figure}[t!] 
   \centering
	\includegraphics[width=0.49\textwidth]{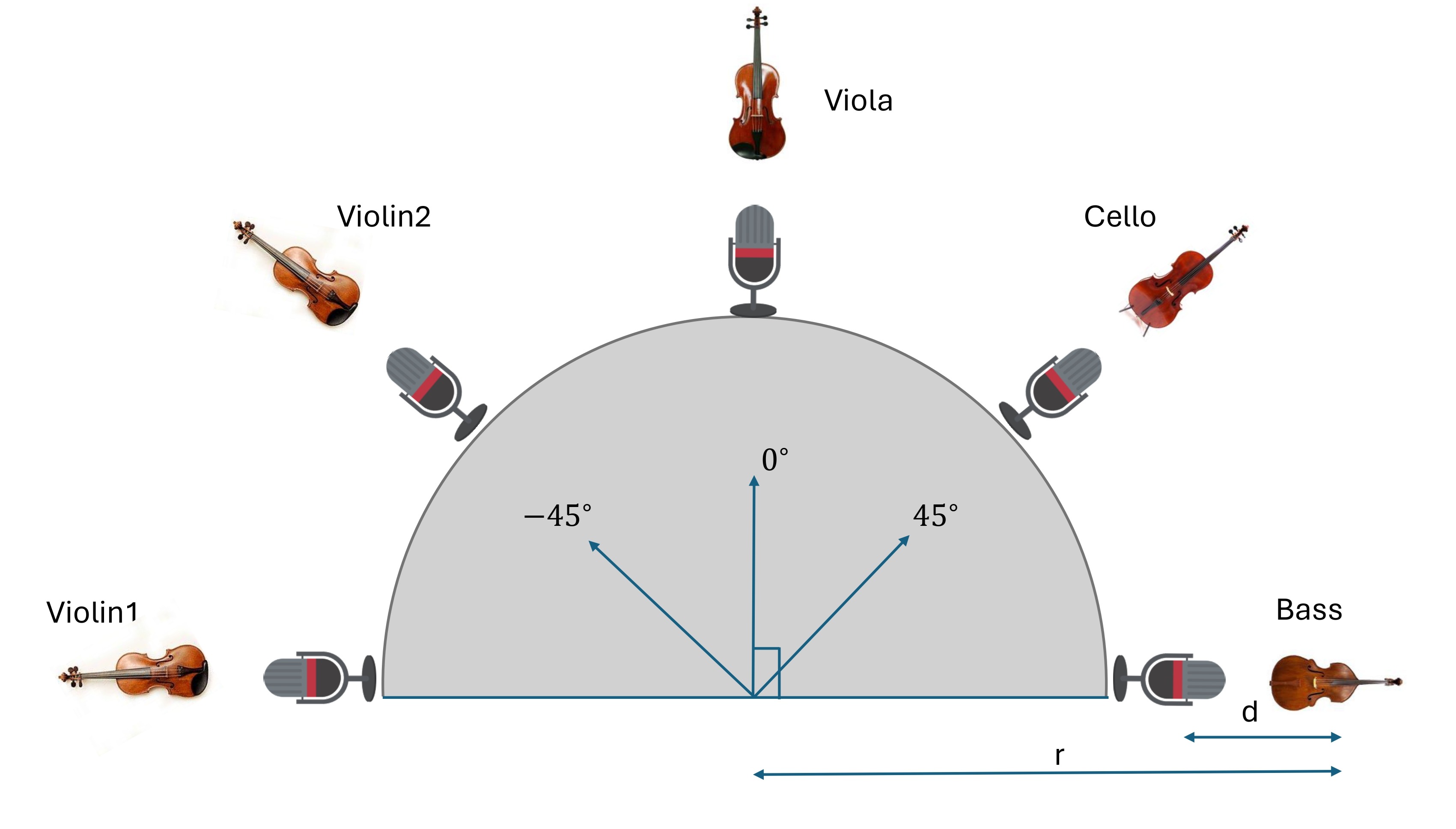}
	\caption{Semicircular string-ensemble layout used in simulations: Violin I/II, Viola, Cello, and Bass positioned from $-90^\circ$ to $90^\circ$ azimuth relative to the Main stereo center; instrument assignments are circularly shifted across runs.}
	\label{fig:inst_position}
\end{figure}

The largest dimension of each simulated room was assigned as its length, and instrument sources were placed relative to a local reference point located at 50\% of the room width and 33\% of its length. Following the typical spatial arrangement of string ensembles, instruments were positioned in a semicircle as shown in Fig. \ref{fig:inst_position}, with corresponding radius and microphone positions given in Table \ref{tab:exp_setup}. Each close microphone is modeled with a cardioid directivity pattern and placed at $d=\SI{25}{cm}$ from its corresponding instrument, oriented toward it unless otherwise specified. We define six ensemble layouts. Layouts L1–L3 are used for training, L4–L5 are reserved for validation and testing, and L6 is used for testing with URMP pieces that do not contain string instruments. It should be noted that, even if the absolute position of each instrument changes across layouts, the assignment of instruments to audio channels remains fixed.
\begin{table}[t!]
\centering
\caption{Room dimensions and RT60 values for the simulated orchestral recording environments. R1, R2, R3, and R4 were used for training, while R5 and R6 were reserved for validation and testing respectively.}
\vspace{6pt}
\label{tab:room_des}
\resizebox{0.99\width}{!}{%
\begin{tabular}{ccc}
\toprule
Room & Size [m] & RT60 [s] \\ 
\midrule
R1 & 36x12x14 & 1.6 \\
R2 & 30x13.9x13 & 1.2 \\
R3 & 17x13x10 & 1.0 \\ 
R4 & 21.5x20x9.2 & 1.2 \\ 
R5 & 20x16x10 & 1.3 \\ 
R6 & 22x19.5x12 & 1.6 \\ 
\bottomrule
\end{tabular}%
}
\end{table}
\begin{table}[t!]
\centering
\caption{Instrument positions used during the experiments. L1, L2, and L3 were used for training, while L4 and L5 were reserved for validation and testing respectively.}
\vspace{6pt}
\label{tab:exp_setup}
\resizebox{0.99\width}{!}{%
\begin{tabular}{lllllll}
\toprule
\multirow{2}{*}{Layout} & \multirow{2}{*}{r [m]} & \multicolumn{5}{c}{Instrument at $\phi$} \\
\cmidrule(lr){3-7} 
& & $-90^\circ$ & $-45^\circ$ & $0^\circ$ & $45^\circ$ & $90^\circ$ \\
\midrule
L1  & 1.5 & Violin1 & Violin2 & Viola & Cello & Bass \\  
L2 & 1.6 & Bass & Violin1 & Violin2 & Viola & Cello \\  
L3 &  1.7 & Cello & Bass & Violin1 & Violin2 & Viola \\
L4 &  1.8 & Viola & Cello & Bass & Violin1 & Violin2 \\ 
L5 & 2 & Violin2 & Viola & Bass & Cello & Violin1 \\ 
L6 & 2 & Inst1 & Inst2 & Inst3 & Inst4 & Inst5 \\ 
\bottomrule
\end{tabular}%
}
\end{table}
 \begin{table}[t!]
    \centering
    \vspace{-6pt}
    \caption{Results of the models evaluated using the string ensembles of URMP as source signals, the room R6, and the L5 instrument layout.}
    \vspace{6pt}
    \label{tab:results}
    \resizebox{0.99\width}{!}{%
    \begin{tabular}{lSS[explicit-sign=+]S[explicit-sign=+]S[explicit-sign=+]S[explicit-sign=+]}
    \toprule
        \multirow{3}{*}{Instrument} & {\multirow{3}{*}{\makecell{Original \\ SDR [dB]}}} & \multicolumn{4}{c}{SDR improvement [dB]} \\
        \cmidrule(lr){3-6} 
        & & \multicolumn{2}{c}{Mag. Spect.} & \multicolumn{2}{c}{R\&I Spect.} \\
        \cmidrule(lr){3-4} \cmidrule(lr){5-6}
        & & {No perm.} & {Perm.} & {No perm.} & {Perm.}    \\
        \midrule 
        Violin 1 & 16.9 &  5.2 & 8.4 & -1.5 & 7.6 \\
        Violin 2 & 15.7 &  4.5 & 7.6 & -2.1 & 6.0 \\
        Viola    & 17.6 &  1.6 & 6.0 & -3.8 & 5.4 \\
        Cello    & 16.2 & -0.8 & 3.1 & -4.3 & 0.9 \\
        Bass     & 18.8 & -8.9 & 1.2 &-16.0 & -0.6 \\
        \midrule
        Average  & 17.0 &  0.3 &  5.3 & -5.5 &  3.9 \\
        \bottomrule

    \end{tabular}%
    }
\end{table}
\subsection{Evaluation metric}
We evaluate the separation performance using the signal-to-distortion ratio (SDR). Following the SiSEC18 evaluation campaign \cite{stoter20182018}, SDR is computed for each instrument on non-overlapping 1-second frames. The aggregated results of every piece are computed by computing the median of its frames, and the aggregated results for the datasets are obtained as the median SDR of their music pieces. The metric is computed without allowing any linear distortion of the ground truth, making it equivalent to the conventional signal-to-noise ratio (SNR).
\begin{table*}[ht!]
    \centering
    \vspace{-6pt}
    \caption{Results of the models evaluated using different source signals and acoustic conditions.}
    \vspace{6pt}
    \label{tab:results2}
    {%
    \begin{tabular}{lllSSS[explicit-sign=+]S[explicit-sign=+]S[explicit-sign=+]S[explicit-sign=+]}
    \toprule
        \multirow{3}{*}{\makecell{Source signals}} & \multirow{3}{*}{\makecell{Room}} & \multirow{3}{*}{\makecell{Inst. layout}} & {\multirow{3}{*}{\makecell{$d$ {[cm]}}}} & {\multirow{3}{*}{\makecell{Original \\ SDR [dB]}}} & \multicolumn{4}{c}{SDR improvement [dB]} \\
        \cmidrule(lr){6-9} 
         & & & & & \multicolumn{2}{c}{Mag. Spect.} & \multicolumn{2}{c}{R\&I Spect.} \\
        \cmidrule(lr){6-7} \cmidrule(lr){8-9}
         & & & & & {No perm.} & {Perm.} & {No perm.} & {Perm.}    \\
        \midrule
        SynthSOD strings & R1 (train) & L2 (train) & 25 & 16.1 &  7.2 &  6.3 &  5.2 &  6.6 \\
        SynthSOD strings & R6 (test)  & L2 (train) & 25 & 16.1 &  6.0 &  6.2 &  3.9 &  6.4 \\
        SynthSOD strings & R6 (test)  & L5 (test)  & 25 & 16.6 &  0.6 &  4.1 & -2.7 &  3.4 \\
        URMP strings     & R6 (test)  & L5 (test)  & 25 & 17.0 &  0.3 &  5.3 & -5.5 &  3.9 \\
        \midrule
        URMP strings     & R6 (test)  & L5 (test)  & 20 & 19.1 & -0.6 &  4.7 & -6.8 &  3.4 \\
        URMP strings     & R6 (test)  & L5 (test)  & 30 & 15.2 &  0.2 &  5.5 & -4.3 &  4.0 \\
        URMP strings     & R6 (test)  & L5 (test) & 40 & 12.7 & -2.1 &  4.9 & -3.2 &  3.8 \\
        URMP strings     & R6 (test)  & L5 (test) & 50 & 10.9 & -0.3 &  4.2 & -2.3 &  3.6 \\
        \midrule
        URMP others      & R6 (test)  & L6 (test) & 25 & 19.0 &  0.8 &  5.7 & -5.5 &  4.2 \\
        \bottomrule

    \end{tabular}%
    }
\end{table*}
\subsection{Results}
Table \ref{tab:results} shows the SDR values obtained by the analyzed models evaluated on the string ensembles of URMP as source signals and the acoustics of room R6 and an unseen instrument layout. The results show that the model trained with magnitude spectrograms and random channel permutations consistently improves SDR in the close microphones of all the instruments. In contrast, the model trained without permutation performs notably poorly, even severely degrading the result of the Bass channel. Interestingly, the models using concatenated real and imaginary parts of the spectrograms as input to the frequency branch of Hybrid Demucs perform worse than those using only magnitude spectrograms. Although the spectrogram phase contains spatial information, we hypothesize that the temporal branch already captures this information. Adding it again in the frequency branch only increases input dimensionality, raising the risk of overfitting without providing additional benefits.
\subsubsection{Generalization analysis}
The first part of Table \ref{tab:results2} shows the results of evaluating the models on datasets that become progressively more different from the training setup. We can see that the models trained without channel permutations perform well when tested on SynthSOD music pieces reserved for testing, as long as the simulated acoustic conditions match those seen during training. Their performance remains comparable when tested in a slightly different room, but collapses completely when the instrument layout changes. In contrast, the models trained with random channel permutations maintain strong performance even under new layouts. The result confirms our hypothesis that permutation-equivariant training encourages the model to infer acoustic conditions from the input audio rather than memorizing them. Finally, the models trained with our proposed channel permutations also generalize well to real recordings from the URMP dataset, demonstrating that they rely primarily on spatial cues rather than the timbral characteristics of individual instruments.
\subsubsection{Impact of the microphone distance}
To evaluate the robustness of the models against mismatches in the distance between the instruments and their close microphones, we tested the models trained with a fixed distance of \SI{25}{cm} on datasets generated with different distances. As shown in the second part of Table \ref{tab:results2}, models trained with random permutation still reduce the microphone bleed of the close microphones even when their distance to the microphone is doubled compared to training. Of course, introducing variability on the micropohone distance during training would probably improve these results, but keeping it constant allowed us to analyze the robustness of the models against unseen conditions. 
\subsubsection{Generalization to unseen instruments}
To further test our hypothesis that permutation-equivariant models do not rely on the learned timbral characteristics of specific instruments, we evaluate the models on URMP pieces containing no string instruments, with each instrument randomly assigned to a channel. As shown in the last part of Table \ref{tab:results2}, the proposed models still achieve strong results in this setting, demonstrating their ability to generalize to unseen instruments.

\section{Conclusion}
\label{sec:concl}
In this paper, we studied cross-instrument bleed in multichannel close-microphone recordings and formulated debleeding as a joint multichannel source separation problem. We introduced a channel-permutation-equivariant training strategy that reduces overfitting to fixed timbres and channel orders. This improves generalization from synthetic training mixtures to unseen instrument layouts, rooms, and real recordings, including cases with unseen instruments. Across our evaluations, permutation-equivariant training consistently improved SDR and reduced cross-talk compared to non-permutation baselines, with the largest gains in the channels most affected by bleed. Overall, permutation equivariance provides a simple, data-centric mechanism to improve robustness under synthetic-to-real and layout mismatches. Future work will evaluate the approach on larger real close-microphone datasets and explore architectures with built-in permutation equivariance to further enhance separation performance.


\bibliographystyle{IEEEtran}
\bibliography{refs}


\end{document}